\documentclass[sigconf]{acmart}

\usepackage{stfloats}

\AtBeginDocument{%
  }

\setcopyright{acmlicensed}
\copyrightyear{2024}
\acmYear{2024}
\acmDOI{XXXXXXX.XXXXXXX}

\acmConference[ESEIW '24]{Make sure to enter the correct
  conference title from your rights confirmation emai}{October 20--25,
  2024}{Barcelona, Spain}
\acmISBN{978-1-4503-XXXX-X/18/06}

\begin{document}

\title{Validation of an Analysability Model in Hybrid Quantum Software}

\author{Ana Díaz-Muñoz}
\orcid{0000-0001-6515-8835}
\email{adiaz@aqclab.es}

\affiliation{%
  \institution{AQCLab Software Quality}
  \institution{University of Castilla-La Mancha}
  \city{Ciudad Real}
  \country{Spain}
}

\author{José A. Cruz-Lemus}
\orcid{0000-0002-0470-609X}
\email{joseantonio.cruz@uclm.es}
\author{Moisés Rodríguez}
\orcid{0000-0003-2155-7409}
\email{moises.rodriguez@uclm.es}
\author{Mario Piattini}
\orcid{0000-0002-7212-8279}
\email{mario.piattini@uclm.es}
\affiliation{%
  \institution{University of Castilla-La Mancha}
  \city{Ciudad Real}
  \country{Spain}
}

\author{Maria Teresa Baldassarre}
\orcid{0000-0001-8589-2850}
\email{mariateresa.baldassarre@uniba.it}
\affiliation{%
  \institution{University of Bari Aldo Moro}
  \city{Bari}
  \country{Italy}
}

\renewcommand{\shortauthors}{Díaz-Muñoz et al.}

\begin{abstract}
In the context of quantum-classical hybrid computing, evaluating analysability, which is the ease of understanding and modifying software, presents significant challenges due to the complexity and novelty of quantum algorithms. Although advances have been made in quantum software development, standard software quality evaluation methods do not fully address the specifics of quantum components, resulting in a gap in the ability to ensure and maintain the quality of hybrid software products. In this registered report proposal, we intend to validate a quality model focused on the analysability of hybrid software through an international collaborative approach involving academic institutions from Italy and Spain through a controlled experiment. This approach allows for a more detailed analysis and validation methodology and establishes a framework for future research and developments in software quality assessment in quantum computing.
\end{abstract}




\maketitle

\section{Problem statement}

In the emerging field of quantum computing, quantum algorithms' integration with conventional software components has led to the development of hybrid software, designed to harness both the superior processing power of quantum computing and the versatility and robustness of classical software \cite{Bernhardt2019}. Therefore, in this work we refer to hybrid software as those applications that combine algorithms written in classical programming languages (e.g., Python) with quantum circuit algorithms, taking advantage of the strengths of both paradigms to solve complex problems more efficiently. This technological fusion can introduce a new paradigm in Software Engineering, which may solve complex problems more efficiently. However, the quality, and more concretely the maintainability of these hybrid software products, as one of its main characteristics, remain a significant challenge \cite{Piattini2020}, especially regarding the analysability of the software.

Analysability, one of the main sub-characteristics of maintainability according to ISO/IEC 25010 \cite{ISO2013}, refers to the ease with which a software product can be understood, diagnosed for deficiencies, and modified for corrections or improvements \cite{Rodrguez2016Evaluation}. In hybrid computing (classical-quantum), analysability is critical due to the inherent complexity and quantum algorithms' novel nature, which poses significant challenges for software developers and maintainers \cite{Akbar2023}.

Despite advances in quantum software development, traditional methods of software quality evaluation \cite{Rodriguez2019} are not fully equipped to address the specifics of quantum components \cite{Murillo2024}, resulting in a gap in organizations' ability to ensure and maintain the quality of their hybrid software products. Our approach focuses on defining metrics at a logical level, independent of the underlying hardware. Although these metrics will eventually vary based on how a circuit is implemented on specific hardware, our method remains hardware-agnostic, emphasizing design rather than physical implementation.

In previous works \cite{Alvarado2023}\cite{Diaz2024SCP}\cite{Diaz2024JUCS}, a quality model focused on the analysability of hybrid software was presented, as well as the automated technological environment that allows measuring and evaluating these algorithms. An analyzability model is a conceptual framework that defines a set of specific properties and metrics for assessing the ease with which a software product can be understood. The main objective of this work is to present a proposal, through a registered report \cite{Ernst2023} to validate this model along with its properties and metrics, aiming to provide a means to measure the analysability of hybrid software. This approach could facilitate better understanding and maintenance of hybrid software \cite{Verdugo2024}, but will also establish a framework for future research and developments in software quality assessment in quantum computing.

To effectively address the challenge of evaluating the analysability of hybrid software, this article is based on a collaborative study and experiment at an international level. We have joined efforts with academic institutions, including the University of Bari (Italy), the University of Extremadura (Spain), the University of Deusto (Spain), and the University of Castilla-La Mancha (Spain). This cross-border collaboration has been crucial in developing a holistic approach that reflects the diverse perspectives and experiences in the field of quantum and classical software engineering but also leverages the diversity of knowledge to enrich the research and validation process. This joint approach has allowed for a rigorous experimental design and exhaustive validation methodology of the proposed model for the analysability of hybrid software. This model has benefited from the contributions of experts in Quantum Software Engineering and software quality assessment, leveraging the latest theoretical advances and best practices in the field. Furthermore, available technological resources and capabilities have supported practical testing and analysis.

This collaborative approach is intended to expand the scope and depth of our study and strengthen the validity and relevance of the results obtained. The interaction between all these universities has facilitated an enriched research environment that has driven the creation of innovative knowledge in an area of growing importance in Quantum Software Engineering. By combining efforts and specialities, we have aimed to overcome limitations in hybrid software evaluation and set a precedent for future collaborations in applied research in quantum computing and hybrid software.

To sum up, the hypothesis presented to carry out the development of this article and the achievement of the final objective is that the proposed analyzability model for hybrid software can potentially identify the level of capability to understand hybrid software systems. This hypothesis will be tested through several controlled experiments with participants from different academic institutions, as described in the following sections.


\section{Methods}

\begin{figure*}[hbt]
    \centering
    \includegraphics[width=0.8\linewidth]{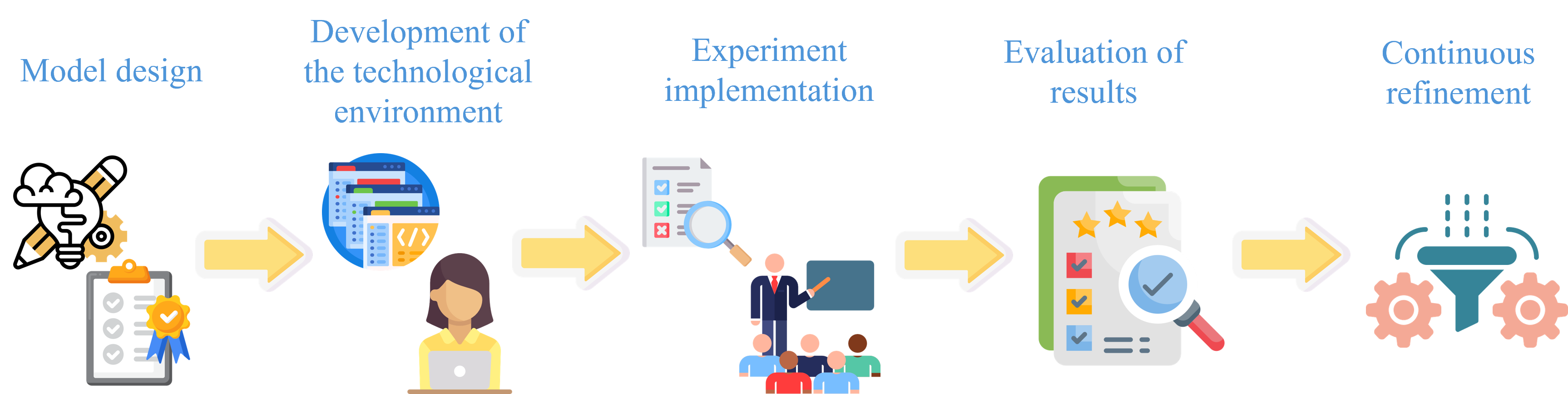}
    \caption{Methodology overview.}
    \label{fig:method}
\end{figure*}

To effectively address the challenge of evaluating and improving the analysability of quantum-classical hybrid software, a structured methodological approach has been designed consisting of several key phases, including model design \cite{Diaz2024JUCS}, technological environment development \cite{Diaz2024SCP}, experiment implementation, and results evaluation (see Fig. \ref{fig:method}). The diagram in Figure \ref{fig:method} includes both the work process and the stages of the research study. This collaborative approach between the University of Bari, the University of Extremadura, the University of Deusto, and the University of Castilla-La Mancha has allowed the integration of different perspectives and technical expertise during validation, strengthening the research and ensuring the robustness of the study.

\subsection{Model design}

The first step in our methodology involves designing a specific analysability model for hybrid software. This model incorporates a set of properties and metrics relevant to classical and quantum components of the software. Several properties influencing analysability were identified, such as code complexity, modularity, documentation, and various elements composing quantum algorithms. To develop this model, a comprehensive review of current research was conducted, and consultations with experts in quantum computing and software engineering were performed \cite{Diaz2024JUCS}.

Our hybrid software quality assessment model incorporates multiple properties and essential components to ensure a comprehensive evaluation. Some of these properties are directly related to the classical components of hybrid software:

\begin{itemize}
    \item Coding Rules: Establishment of standards and best practices for writing classical and quantum code to ensure software readability and maintainability.
    \item Code Documentation: Evaluation of the quality and thoroughness of the documentation, ensuring the code is well-commented and explained to facilitate understanding by other developers.
    \item Cyclomatic Complexity: Measurement of the control flow complexity of classical code, helping to identify areas of the software that may be prone to errors and difficult to maintain.
    \item Package and Class Structuring: Analysis of the organization of the code into packages and classes, evaluating cohesion and coupling to ensure a modular and scalable design.
    \item Method Size: Evaluation of the size of methods in terms of lines of code and functional complexity, promoting concise and specific methods.
    \item Duplicate Code: Identification and quantification of duplicate code to reduce redundancy and facilitate maintenance.
 \end{itemize}

On the other hand, there are also some other properties related to the quantum circuit components:
\begin{itemize}

    \item Circuit Width: Number of qubits used in the quantum circuit, indicating the scale of the implemented quantum system.
    \item Circuit Depth: Number of layers of operations in the circuit, related to the duration and decoherence in a quantum system.
    \item Gate Complexity: Evaluation of the number and type of quantum gates used, indicating the operational complexity of the circuit.
    \item Conditional Instructions: Analysis of the presence and use of conditional operations in the quantum code.
    \item Quantum Cyclomatic Complexity: Measurement of the control flow complexity in quantum algorithms.
    \item Measurement Operations: Evaluation of the quantity and positioning of measurement operations in the circuit.
    \item Initialization and Reset Operations: Analysis of how and when qubits are initialized and reset during program execution.
    \item Auxiliary Qubits: Identification and use of auxiliary qubits for intermediate operations and their impact on circuit efficiency.
\end{itemize}

Additionally, several implementations of the metrics and corresponding calculation methods were performed to evaluate each of the quantum software's properties. An example of a calculation method for one of the quantum properties is detailed below:

\textbf{How is circuit width calculated?}

In the case of the circuit width, once the different circuits of the product under evaluation have been classified according to the number of qubits they contain, their distribution is evaluated using the following base metrics, with level 3 being the lowest severity level and level 1 being the highest:
\begin{itemize}
    \item \textbf{NC\_CW1}: Number of circuits with level 1 circuit width.
    \item \textbf{NC\_CW2}: Number of circuits with level 2 circuit width.
    \item \textbf{NC\_CW3}: Number of circuits with level 3 circuit width.
    \item \textbf{NCIR}: Total number of circuits.
\end{itemize}

The calculation of the base metrics defined for the circuit width property is obtained from the thresholds shown in Table~\ref{tab:circuit-width-levels}, which are based on the experience gained from various measurements. This table divides the circuits from best to worst depending on the number of qubits contained in each circuit.

\begin{table}[htb]
    \centering
    \begin{tabular}{|c|c|c|}
        \hline
        Levels & Ranges & Quality levels \\
        \hline
        1 & $>$ 15 & Bad circuit width \\
        2 & [9-15] & Regular circuit width \\
        3 & [1-8] & Good circuit width \\
        \hline
    \end{tabular}
    \caption{Definition of the levels for the evaluation of circuit width}
    \label{tab:circuit-width-levels}
\end{table}

From the above base metrics, the following derived metrics are calculated:
\begin{itemize}
    \item \textbf{DC\_CW1}: Circuit density with level 1 circuit width.
    \item \textbf{DC\_CW2}: Circuit density with level 2 circuit width.
    \item \textbf{DC\_CW3}: Circuit density with level 3 circuit width.
\end{itemize}
These derived metrics are calculated by dividing the number of circuits in each level by the total number of circuits evaluated. Equation~\ref{eq:DC_CW1} shows how the derived metric of circuit density with level 1 circuit width is obtained. The remaining derived metrics used in the evaluation of this property are calculated similarly.

\begin{equation}
\text{DC\_CW1} = \frac{\text{NC\_CW1}}{\text{NCIR}}
\label{eq:DC_CW1}
\end{equation}

Once the three densities have been obtained, a standardized quality value for the property is obtained using the profile function. In the profile function for circuit width, the following ranges are defined (see Table~\ref{tab:profile-function}). For each quality level, the maximum accepted threshold of circuit density is indicated. The table does not include the most desirable width level, which corresponds to good circuit width.

\begin{table}[htb]
    \centering
    \begin{tabular}{|c|c|c|c|c|}
        \hline
        Ranges & Level 1 & Level 2 & Quality Levels \\
        \hline
        0 & - & - & 0 \\
        \hline
        1 & 20 & 40 & [0-33) \\
        \hline
        2 & 15 & 30 & [33-66) \\
        \hline
        3 & 10 & 20 & [66-100) \\
        \hline
        4 & 7 & 15 & 100 \\
        \hline
    \end{tabular}
    \caption{Definition of the ranges for the evaluation of circuit width}
    \label{tab:profile-function}
\end{table}

The analysability model was developed through a structured process. Initially, we investigated the fundamentals of classical and quantum computing and identified specific properties and requirements of hybrid software. We then designed a modular model structure to measure both classical and quantum components and implemented algorithms for evaluation. Subsequently, we identified and defined appropriate metrics to evaluate hybrid software's analysability, considering the model's properties and components. This approach was designed to ensure the robustness and effectiveness of the model before its experimental evaluation.

\subsection{Development of the technological environment}

\begin{figure*}[htb]
    \centering
    \includegraphics[width=0.55\linewidth]{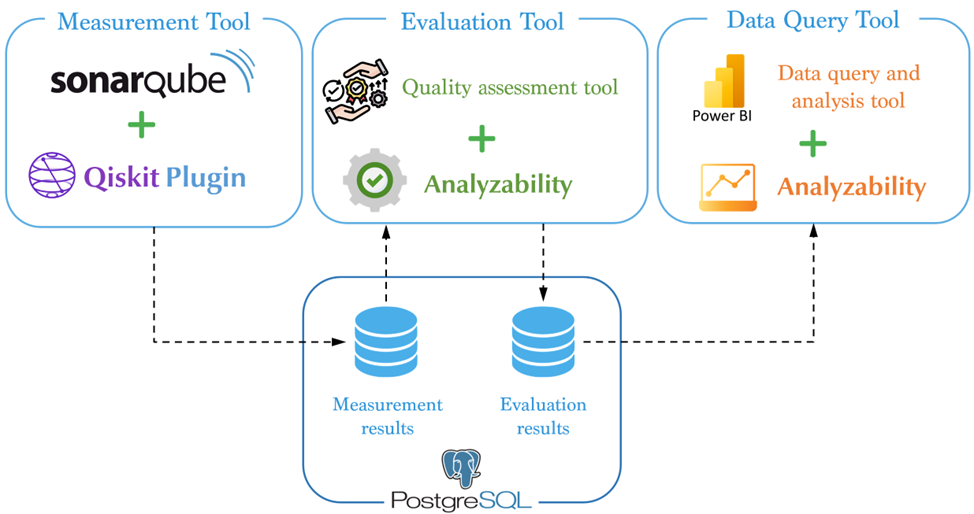}
    \caption{Technological environment.}
    \label{fig:architecture}
\end{figure*}

\begin{figure*}[htb]
    \centering
    \includegraphics[width=0.55\linewidth]{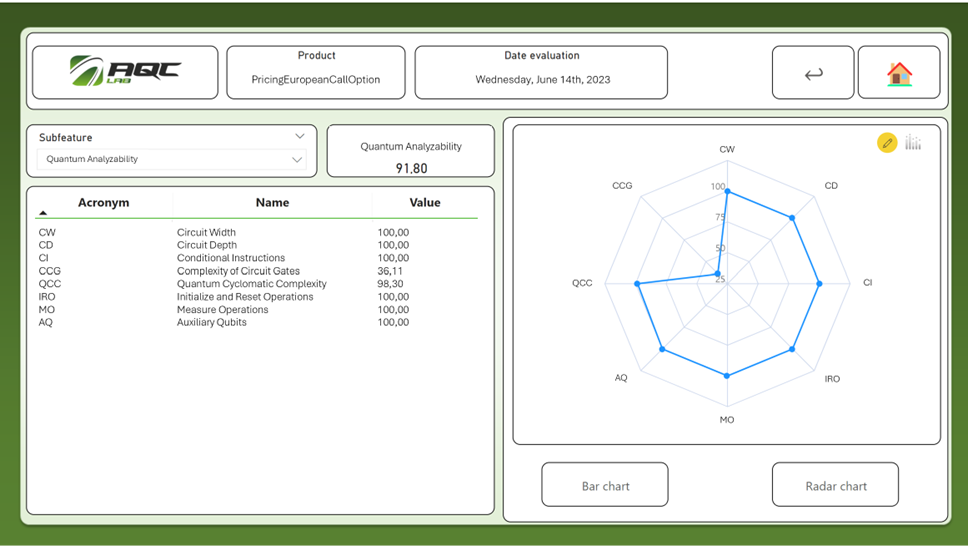}
    \caption{Results visualization tool.}
    \label{fig:powerbi}
\end{figure*}

Subsequently, a technological environment was developed to enable the implementation of the model and the evaluation of hybrid software. This environment includes emerging tools such as a plug-in for SonarQube developed for the automatic measurement of the metrics defined in the model (see Fig. \ref{fig:architecture}) and a platform designed in Power BI (see Fig. \ref{fig:powerbi}) to visualize the results of these measurements and the final evaluation \cite{Diaz2024SCP}.

The technological environment developed in our work has been preliminarily validated by measuring and evaluating recognized quantum algorithms implemented using Qiskit (\url{https://www.ibm.com/quantum/qiskit}) and Quirk (\url{https://algassert.com/quirk}) \cite{Alvarado2023}. The ability to apply our environment to algorithms widely used in the quantum community suggests the relevance and practical applicability of our research in real software development environments.

\section{Experiments design and implementation}

With the developed model and technological environment in place, an experiment has been designed and implemented to validate the applicability and effectiveness of the analysability model. This experiment intends to evaluate the analysability of hybrid software by applying several metrics on different quantum algorithms developed specifically for this purpose. Real case studies and available technologies such as the Quirk and Qiskit frameworks will be used to ensure that the analysis and results are practical and relevant for hybrid software developers.

The implementation of this experiment has been structured into several crucial stages, starting with intensive training sessions on quantum computing, followed by the practical application of this training in a controlled experimental environment. These stages will be described further next.

\subsection{Experimental hypothesis}

As previously commented, this empirical study hypothesises that the proposed analyzability model for hybrid software can potentially identify the level of capability to understand hybrid software systems. This hypothesis will be tested through several controlled experiments with participants from different academic institutions, as described in the following sections.

This hypothesis will be tested through a series of controlled experiments with participants from different academic institutions, as described in the following sections.

\subsubsection{Experiment methodology}

The experiment will be carried out in May 2024 at the University of Bari with two sets of students: undergraduate (third year) students with a BSc in Computer Science and a Master's Degree in Computer Science students.

The experiment will begin with several training sessions designed to equip the subjects with essential knowledge about quantum software and the tools and environments to be used. We anticipate involving approximately 100 students and 15 researchers in our experiments. This sample size is chosen to ensure a representative dataset for evaluating the analysability model. The training agenda will be distributed as follows:

\begin{itemize}
    \item \textbf{Session I.} Training session conducted by José A. Cruz-Lemus. It will cover the fundamentals of Quantum Software Engineering (QSE) and quantum circuits, together with Quirk as a simulation tool. The session will last approximately 2-2,5 hours.
        
    \item \textbf{Session II.} Training session conducted by Ana Díaz-Muñoz, focusing on the fundamentals of quantum software development plus an overview of Qiskit as a quantum software development language. Some examples of the experimental tasks will also be performed. The session will last approximately 2-2,5 hours.

    \item \textbf{Session III.} Experimental session in which the analysability model for quantum software and circuits will be evaluated. The session will last approximately 1 hour.
\end{itemize}

Session I will take place one day, while sessions II and III will be done on a different day.

Later, in mid-June, a replication of the experiment will be held online involving subjects from the universities of Extremadura, Deusto, and Castilla-La Mancha. In this case, the subjects will be quantum computing researchers, so they will need reduced training sessions. Later they will perform the experimental session using the same materials as the students. This replication will allow us to validate the preliminary obtained results and ensure the consistency of the analysability model in different contexts and user groups.

In this work, we focus on the validation of the quantum software part, since the classical part of the system is currently validated by the AQCLab laboratory (\url{https://aqclab.es/}), which is accredited by ILAC (\url{https://ilac.org/}) and ENAC (\url{https://www.enac.es/}) for the evaluation of traditional software under the ISO 25000 standard \cite{ISO2014} \cite{Rodriguez2015}. This means that the classical software has been tested and certified to meet the established standards and requirements. By focusing on the validation of the model for the quantum part, we seek to ensure that the quantum software works correctly and efficiently, which is fundamental for the correct functioning of the hybrid system as a whole.

\subsubsection{Analyzability model validation testing}

The materials for validating the analysability model for quantum software and circuits have been designed to assess the effectiveness of the proposed model in the real context of quantum software development and analysis. The process begins with the collection of demographic data from the experimental subjects, which will help contextualize the experiment results and ensure that the conclusions are representative and generalisable to different demographic groups. All the forms will be accessed online by the subjects using Google Forms. Their main details are explained next:

\begin{itemize}
    \item \textbf{Demographic questionnaire.} Before going into the technical forms, each subject will be given a demographic questionnaire. This questionnaire has been designed to capture a detailed profile of the participants, including various aspects such as gender, age, geographic location, educational level, primary language spoken at home, and experience in both quantum and classical computing. The collection of demographic information in our study serves key objectives, such as contextualizing the results, and allowing adjustments in the analysis based on specific demographic variables such as age, gender, and previous experience. This will help understand how these factors may influence participants' ability to use and analyze quantum software. Additionally, it ensures that the study group is diverse and representative of the population targeted by the analysability model. It will also facilitate conducting statistical analyses that may uncover correlations between demographic characteristics and performance in analysability tasks, thus providing valuable insights for improving the model and related tools.

    \item \textbf{Experimental Tasks.} The experimental questionnaires designed for this study are crucial for validating the analys\-ability model in the context of quantum computing. These consist of five different questionnaires, each associated with a circuit reflecting one of the five levels of analysability defined in the model. Each participant will also be provided with the corresponding quantum algorithm for a circuit (written using Qiskit). The task to be performed is to transform the algorithm into the appropriate quantum circuit (on Quirk). Later, the subject must answer a total of eight questions for each circuit, focused on the values obtained for different specific quantum states. This methodology allows for an effective and detailed assessment of users' ability to understand and analyze quantum circuits and algorithms, thus contributing to the validation and improvement of the various properties comprising the designed analysability model. Each subject will receive the same five forms but in a randomized order. This way, the threats of the tiring and learning effects will be alleviated.
    
\end{itemize}

\subsection{Evaluation of results}

The final phase of the methodology involves evaluating the results obtained from the experiments. The collected data will be analyzed to determine the model's effectiveness in assessing the analysability of hybrid software.

The dataset will include demographic information about the participants, encompassing both students and professionals. This demographic information will be crucial in our data analysis to ensure a comprehensive understanding of the results and to account for potential differences between these groups.

\subsection{Continuous refinement}

Based on the initial evaluation, the model and associated tools are subject to a process of continuous refinement and improvement. This way, it will be possible to adapt to the rapid changes in quantum technology and software development practices.

\section{Results}

The results phase of our research will focus on analyzing the effectiveness of the developed analysability model through the transformation and evaluation of quantum algorithms into circuits, conducted by the study subjects. We expect that the data collected through the experimental questionnaires will provide a solid foundation for assessing the accuracy and applicability of our model in a real-world environment.

\subsection{Quantitative analysis}

The quantitative results stem from the participants' performance in the questionnaires related to each of the five levels of analysability. The success rate in correctly transforming algorithms into quantum circuits and the accuracy in answering specific questions about quantum states will be calculated.

\subsection{Correlation with demographic data}

The statistical analysis of the collected demographic data will help to correlate participants' characteristics, such as age, gender, education, and previous experience in quantum computing, with their performance in the experiment. This analysis will help to highlight any significant trends or deviations that could influence the effectiveness of the analysability model.

\subsection{Preliminary conclusions}

We intend that the preliminary conclusions indicate whether the analysability model is effective when distinguishing between different levels of complexity in both software and quantum circuits and whether it can be a valuable tool for quality evaluators and developers to assess and improve their understanding of these advanced concepts. However, areas for improvement can also be identified. We consider these results to be crucial for the ongoing validation and refinement of the analysability model, ensuring that it is robust, applicable, and effective in enhancing the quality of quantum software.

\section{Significance}

This proposal aims to make significant contributions to the emerging field of Quantum Computing, focusing especially on the integration of quantum algorithms with conventional software components to create hybrid software systems. The main objective of our work is to address a significant gap in the current landscape of software engineering: the analysability of hybrid software, which combines the paradigms of classical and quantum computing. This proposal, once fully explained and written, could be considered interesting and relevant for publication in a scientific and professional knowledge journal. Its importance and value will be stated in the following sub-sections.

\subsection{Main contributions}

\begin{itemize}
    \item \textbf{Development of a specialized analysability model.} The first notable contribution is the creation of a custom analysability model for hybrid software. This model is innovative in its comprehensive approach to incorporating classical and quantum components, taking into account factors such as code complexity, modularity, and documentation, as well as the various elements of circuits, crucial for effective software maintenance and improvement. The model was developed through rigorous research in the field, aiming to reflect the latest in theoretical and practical aspects of Quantum Computing and Software Engineering. This model was initially developed in our previous works and forms the foundation upon which our current study builds \cite{Diaz2024JUCS}.
    
    \item \textbf{Creation and implementation of a technological environment.} Another intended contribution is the development of a technological environment designed to implement our analysability model. This environment features cutting-edge tools such as SonarQube for measuring various metrics and properties, and Power BI for visualizing the final evaluation results. This setup not only facilitates accurate assessment of hybrid software but also enhances the accessibility and interpretability of the analysis results, making it a valuable tool for software developers or maintainers, as well as researchers. The creation and initial implementation of this environment was also part of our prior work \cite{Diaz2024SCP}.

    \item \textbf{Empirical validation through international collaboration.} Finally, and directly related to this proposal, our work aims to be distinguished by its effectiveness and reliability, pending validation through an international collaboration experiment involving several prestigious institutions. This collaborative effort allows us to refine our methodologies and ensure the robustness and applicability of the model in different contexts and user groups. This empirical validation through international collaboration is the primary contribution of the current study, providing new insights and refinement to the previously developed model and technological environment.

\end{itemize}

\subsection{Importance of our work}

This work is aimed to establish a new benchmark for quality and maintainability in the rapidly evolving domain of quantum computing and hybrid software development, excelling in the following aspects:

\begin{itemize}
    \item \textbf{Improve software quality.} Enhanced analysability facilitates maintenance and reduces the number of defects, thereby improving the overall quality and reliability of hybrid software products.
    \item \textbf{Facilitate future research and development.} Our model and methodology provide a foundational framework upon which other researchers can build, driving further innovations in Quantum Software Engineering.
    \item \textbf{Practical application support.} While quantum supremacy has not yet been fully achieved, our research provides practical applications for the current and near-future landscape of quantum computing. These tools bridge the gap between quantum theory and real-world software development, preparing practitioners for future hardware advancements. Our experiments demonstrate how focusing on analysability can effectively maintain hybrid software, paving the way for future developments.
\end{itemize}

\section*{Acknowledgements}

This research was supported by the projects QSERV: Quantum Service Engineering: Development Quality, Testing \& Security of Quantum Microservices (PID2021-124054OB-C32), funded by the Spanish Ministry of Science and Innovation and ERDF; Q2SM: Quality Quantum Software Model (13/22/IN/032) project financed by the Junta de Comunidades de Castilla-La Mancha and FEDER funds; Financial support for the execution of applied research projects, within the framework of the UCLM Own Research Plan, co-financed at 85\% by the European Regional Development Fund (FEDER) UNION (2022-GRIN-34110); and AETHER-UCLM: A holistic approach to smart data for context-guided data analysis (PID2020-112540RB-C42) funded by MCIN/AEI/10.13039/501100011033/. It also received financial support for the execution of applied research projects, within the framework of the UCLM Own Research Plan, co-financed at 85\% by the European Regional Development Fund (FEDER) UNION (2022-GRIN-34110).

\bibliographystyle{ACM-Reference-Format}
\bibliography{sample-base}


\begin{thebibliography}{14}


\ifx \showCODEN    \undefined \def \showCODEN     #1{\unskip}     \fi
\ifx \showDOI      \undefined \def \showDOI       #1{#1}\fi
\ifx \showISBNx    \undefined \def \showISBNx     #1{\unskip}     \fi
\ifx \showISBNxiii \undefined \def \showISBNxiii  #1{\unskip}     \fi
\ifx \showISSN     \undefined \def \showISSN      #1{\unskip}     \fi
\ifx \showLCCN     \undefined \def \showLCCN      #1{\unskip}     \fi
\ifx \shownote     \undefined \def \shownote      #1{#1}          \fi
\ifx \showarticletitle \undefined \def \showarticletitle #1{#1}   \fi
\ifx \showURL      \undefined \def \showURL       {\relax}        \fi
\providecommand\bibfield[2]{#2}
\providecommand\bibinfo[2]{#2}
\providecommand\natexlab[1]{#1}
\providecommand\showeprint[2][]{arXiv:#2}

\bibitem[Akbar et~al\mbox{.}(2023)]%
        {Akbar2023}
\bibfield{author}{\bibinfo{person}{Muhammad~Azeem Akbar}, \bibinfo{person}{Arif~Ali Khan}, {and} \bibinfo{person}{Saima Rafi}.} \bibinfo{year}{2023}\natexlab{}.
\newblock \showarticletitle{A systematic decision-making framework for tackling quantum software engineering challenges}.
\newblock \bibinfo{journal}{\emph{Automated Software Engg.}} \bibinfo{volume}{30}, \bibinfo{number}{2} (\bibinfo{date}{jul} \bibinfo{year}{2023}), \bibinfo{numpages}{44}~pages.
\newblock
\showISSN{0928-8910}
\urldef\tempurl%
\url{https://doi.org/10.1007/s10515-023-00389-7}
\showDOI{\tempurl}


\bibitem[Alvarado-Valiente et~al\mbox{.}(2023)]%
        {Alvarado2023}
\bibfield{author}{\bibinfo{person}{Jaime Alvarado-Valiente}, \bibinfo{person}{Javier Romero-{\'A}lvarez}, \bibinfo{person}{Ana D{\'i}az}, \bibinfo{person}{Mois{\'e}s Rodr{\'i}guez}, \bibinfo{person}{Ignacio Garc{\'i}a-Rodr{\'i}guez}, \bibinfo{person}{Enrique Moguel}, \bibinfo{person}{Jose Garcia-Alonso}, {and} \bibinfo{person}{Juan~M. Murillo}.} \bibinfo{year}{2023}\natexlab{}.
\newblock \showarticletitle{Quantum Services Generation and Deployment Process: A Quality-Oriented Approach}. In \bibinfo{booktitle}{\emph{Quality of Information and Communications Technology}}, \bibfield{editor}{\bibinfo{person}{Jos{\'e}~Maria Fernandes}, \bibinfo{person}{Guilherme~H. Travassos}, \bibinfo{person}{Valentina Lenarduzzi}, {and} \bibinfo{person}{Xiaozhou Li}} (Eds.). \bibinfo{publisher}{Springer Nature Switzerland}, \bibinfo{address}{Cham}, \bibinfo{pages}{200--214}.
\newblock
\showISBNx{978-3-031-43703-8}


\bibitem[Bernhardt(2019)]%
        {Bernhardt2019}
\bibfield{author}{\bibinfo{person}{Chris Bernhardt}.} \bibinfo{year}{2019}\natexlab{}.
\newblock \bibinfo{booktitle}{\emph{Quantum Computing for Everyone}}.
\newblock
\showISBNx{9780262350914}
\urldef\tempurl%
\url{https://doi.org/10.7551/mitpress/11860.001.0001}
\showDOI{\tempurl}


\bibitem[Díaz-Muñoz et~al\mbox{.}(2024)]%
        {Diaz2024SCP}
\bibfield{author}{\bibinfo{person}{Ana Díaz-Muñoz}, \bibinfo{person}{Moisés Rodríguez}, {and} \bibinfo{person}{Mario Piattini}.} \bibinfo{year}{2024}\natexlab{}.
\newblock \showarticletitle{Implementing an environment for hybrid software evaluation}.
\newblock \bibinfo{journal}{\emph{Science of Computer Programming}}  \bibinfo{volume}{236} (\bibinfo{year}{2024}), \bibinfo{pages}{103109}.
\newblock
\showISSN{0167-6423}
\urldef\tempurl%
\url{https://doi.org/10.1016/j.scico.2024.103109}
\showDOI{\tempurl}


\bibitem[Ernst and Baldassarre(2023)]%
        {Ernst2023}
\bibfield{author}{\bibinfo{person}{Neil~A. Ernst} {and} \bibinfo{person}{Maria~Teresa Baldassarre}.} \bibinfo{year}{2023}\natexlab{}.
\newblock \showarticletitle{Registered reports in software engineering}.
\newblock \bibinfo{journal}{\emph{Empirical Software Engineering}} \bibinfo{volume}{28}, \bibinfo{number}{2} (\bibinfo{year}{2023}).
\newblock
\urldef\tempurl%
\url{https://doi.org/10.1007/s10664-022-10277-5}
\showDOI{\tempurl}


\bibitem[ISO/IEC(2011)]%
        {ISO2013}
\bibfield{author}{\bibinfo{person}{ISO/IEC}.} \bibinfo{year}{2011}\natexlab{}.
\newblock \showarticletitle{ISO / IEC 25010 : 2011 Systems and software engineering — Systems and software Quality Requirements and Evaluation ( SQuaRE ) — System and software quality models}.
\newblock
\urldef\tempurl%
\url{https://api.semanticscholar.org/CorpusID:19022091}
\showURL{%
\tempurl}


\bibitem[ISO/IEC(2014)]%
        {ISO2014}
\bibfield{author}{\bibinfo{person}{ISO/IEC}.} \bibinfo{year}{2014}\natexlab{}.
\newblock \showarticletitle{Systems and software engineering — Systems and software Quality Requirements and Evaluation (SQuaRE) — Guide to SQuaRE}.
\newblock
\urldef\tempurl%
\url{https://www.iso.org/standard/64764.html}
\showURL{%
\tempurl}


\bibitem[Murillo et~al\mbox{.}(2024)]%
        {Murillo2024}
\bibfield{author}{\bibinfo{person}{Juan~M. Murillo} {et~al\mbox{.}}} \bibinfo{year}{2024}\natexlab{}.
\newblock \showarticletitle{{Challenges of Quantum Software Engineering for the Next Decade: The Road Ahead}}.
\newblock  (\bibinfo{date}{4} \bibinfo{year}{2024}).
\newblock
\showeprint[arxiv]{2404.06825}~[cs.SE]


\bibitem[Muñoz et~al\mbox{.}(2024)]%
        {Diaz2024JUCS}
\bibfield{author}{\bibinfo{person}{Ana~Díaz Muñoz}, \bibinfo{person}{Moisés~Rodríguez Monje}, {and} \bibinfo{person}{Mario Gerardo~Piattini Velthuis}.} \bibinfo{year}{2024}\natexlab{}.
\newblock \showarticletitle{Towards a set of metrics for hybrid (quantum/classical) systems maintainability}.
\newblock \bibinfo{journal}{\emph{JUCS - Journal of Universal Computer Science}} \bibinfo{volume}{30}, \bibinfo{number}{1} (\bibinfo{year}{2024}), \bibinfo{pages}{25--48}.
\newblock
\showISSN{0948-695X}
\urldef\tempurl%
\url{https://doi.org/10.3897/jucs.99348}
\showDOI{\tempurl}
\showeprint{https://doi.org/10.3897/jucs.99348}


\bibitem[Piattini et~al\mbox{.}(2020)]%
        {Piattini2020}
\bibfield{author}{\bibinfo{person}{Mario Piattini}, \bibinfo{person}{Guido Peterssen~Nodarse}, \bibinfo{person}{Ricardo Pérez-Castillo}, \bibinfo{person}{Jose~Luis Hevia~Oliver}, \bibinfo{person}{Manuel Serrano}, \bibinfo{person}{Guillermo Hernández~González}, \bibinfo{person}{Ignacio Guzmán}, \bibinfo{person}{Claudio Andrés~Paradela}, \bibinfo{person}{Macario Polo}, \bibinfo{person}{Ezequiel Murina}, \bibinfo{person}{Luis Jiménez~Navajas}, \bibinfo{person}{Juan Marqueño}, \bibinfo{person}{Ramses Gallego}, \bibinfo{person}{Jordi Tura}, \bibinfo{person}{Frank Phillipson}, \bibinfo{person}{Juan Murillo}, \bibinfo{person}{Alfonso Niño}, {and} \bibinfo{person}{Moisés Rodríguez}.} \bibinfo{year}{2020}\natexlab{}.
\newblock \showarticletitle{The Talavera Manifesto for Quantum Software Engineering and Programming}.
\newblock  (\bibinfo{date}{03} \bibinfo{year}{2020}).
\newblock


\bibitem[Rodr{\'i}guez et~al\mbox{.}(2016)]%
        {Rodrguez2016Evaluation}
\bibfield{author}{\bibinfo{person}{Mois{\'e}s Rodr{\'i}guez}, \bibinfo{person}{Jes{\'u}s~Ramon Oviedo}, {and} \bibinfo{person}{Mario~G. Piattini}.} \bibinfo{year}{2016}\natexlab{}.
\newblock \showarticletitle{Evaluation of Software Product Functional Suitability: A Case Study}.
\newblock \bibinfo{journal}{\emph{Software Quality Professional Magazine}}  \bibinfo{volume}{18} (\bibinfo{year}{2016}).
\newblock
\urldef\tempurl%
\url{https://api.semanticscholar.org/CorpusID:114609549}
\showURL{%
\tempurl}


\bibitem[Rodriguez et~al\mbox{.}(2019)]%
        {Rodriguez2019}
\bibfield{author}{\bibinfo{person}{Moises Rodriguez}, \bibinfo{person}{Mario Piattini}, {and} \bibinfo{person}{Christof Ebert}.} \bibinfo{year}{2019}\natexlab{}.
\newblock \showarticletitle{Software Verification and Validation Technologies and Tools}.
\newblock \bibinfo{journal}{\emph{IEEE Software}} \bibinfo{volume}{36}, \bibinfo{number}{2} (\bibinfo{year}{2019}), \bibinfo{pages}{13--24}.
\newblock
\urldef\tempurl%
\url{https://doi.org/10.1109/MS.2018.2883354}
\showDOI{\tempurl}


\bibitem[Rodríguez et~al\mbox{.}(2015)]%
        {Rodriguez2015}
\bibfield{author}{\bibinfo{person}{Moisés Rodríguez}, \bibinfo{person}{Mario Piattini}, {and} \bibinfo{person}{C.M. Fernandez}.} \bibinfo{year}{2015}\natexlab{}.
\newblock \showarticletitle{A hard look at software quality: Pilot program uses ISO/IEC 25000 family to evaluate, improve and certify software products}.
\newblock   \bibinfo{volume}{48} (\bibinfo{date}{09} \bibinfo{year}{2015}), \bibinfo{pages}{30--36}.
\newblock


\bibitem[Verdugo et~al\mbox{.}(2024)]%
        {Verdugo2024}
\bibfield{author}{\bibinfo{person}{J. Verdugo}, \bibinfo{person}{J. Oviedo}, \bibinfo{person}{M. Rodriguez}, {and} \bibinfo{person}{M. Piattini}.} \bibinfo{year}{2024}\natexlab{}.
\newblock \showarticletitle{Connecting Research and Practice for Software Product Quality Evaluation and Certification: A Software Laboratory’s 25-Year Journey}.
\newblock \bibinfo{journal}{\emph{IEEE Software}} \bibinfo{volume}{41}, \bibinfo{number}{03} (\bibinfo{date}{may} \bibinfo{year}{2024}), \bibinfo{pages}{33--40}.
\newblock
\showISSN{1937-4194}
\urldef\tempurl%
\url{https://doi.org/10.1109/MS.2024.3357119}
\showDOI{\tempurl}


\end{thebibliography}

\end{document}